# Self-Induced Valley Bosonic Stimulation of Exciton-Polaritons in a Monolayer Semiconductor

Pasquale Cilibrizzi[1], Xiaoze Liu[1], Peiyao Zhang[1], Chenzhe Wang[1],

Quanwei Li[1], Sui Yang[1], Xiang Zhang[1]

[1] NSF Nanoscale Science and Engineering Center, University of California, Berkeley, California 94720, USA

The newly discovered valley degree of freedom in atomically thin two-dimensional transition metal dichalcogenides (TMD) offers a promising platform to explore rich nonlinear physics, such as spinor Bose-Einstein condensate (BEC) and novel valleytronics applications. However, the critical nonlinear effect, such as valley polariton bosonic stimulation, has long remained an unresolved challenge due to the generation of limited polariton ground state densities necessary to induce the stimulated scattering of polaritons in specific valleys. Here, we report the self-induced valley bosonic stimulation of exciton-polaritons via spin-valley locking in a $WS_2$ monolayer microcavity. This is achieved by the resonant injection of valley polaritons at specific energy and wavevector, which allows spin-polarized polaritons to efficiently populate their ground state and induce a valley-dependent bosonic stimulation. As a result, we observe the nonlinear self-amplification of polariton emission from the valley-dependent ground state. Our finding paves the way for the investigation of spin ordering and phase transitions in TMD polariton BEC, offering a promising route for the realization of polariton spin lattices in moiré polariton systems and spin-lasers.

The optical properties of a semiconductor are intrinsically related to its electronic band structure. The broken inversion symmetry in recently emerging two-dimensional transition metal dichalcogenides (TMDs) leads to two energy degenerate valleys at two corners of the first Brillouin zone, referred to as K/K' valleys [1,2]. Strong spin-orbit coupling locks the valley degree of freedom to the electron and hole spins (i.e., spin-valley locking), resulting in specific optical selection rules of excitons [3]. Specifically, excitons in K/K' valleys can be selectively addressed using $\sigma+/\sigma-$ circularly polarized optical excitations. These intriguing properties and the possibility to control valley degree of freedom open new ways to explore spinor/valleytronic physics and novel valleytronics applications [4].

TMD exciton-polaritons (hereafter polaritons), resulting from the strong coupling of cavity photons with TMD excitons [5], are composite bosonic quasi-particles, which can also be optically initialized in specific valleys [6-10]. Recently, TMD polaritons have emerged as promising candidates for studying fundamental polaritonic physics and nonlinearities [11-13]. Different polariton nonlinear behaviours have been explored so far, leading to the observation of polariton condensation [11,12] and moiré nonlinearities [13]. Both these observations result from polariton-polariton interactions inherited by the underlying TMD excitons' Coulomb interactions. Furthermore, polaritons interactions are spin dependent [14]. The spin interactions enrich the physics of polariton condensates, giving rise to many fascinating topological effects in quantum fluids [15]. In this regard, the valley degree of freedom represents an additional experimental knob to control polariton spin interactions in the nonlinear regime. It offers the possibility to engineer polariton spin interactions in microcavities and control stochastic spin phenomena in spinor BEC [16], allowing, for example, the realization of different spin configurations when the polarization of the polariton condensates is pinned to a specific crystallographic axis of the microcavity [17,18]. The possibility to control ferromagnetic spin interactions in polariton condensates, provides new opportunities to experimentally investigate the complex interplays between magnetic order and superfluidity in polariton BECs [19] and explore spin order and phase transitions in moiré polariton systems [13]. However, so far, the valley degree of freedom has been mainly studied in the linear regime. In such a regime, the most fascinating polaritonic physics, such as BEC and superfluidity [14], is precluded.

A critical step to advance TMD polaritons, as a practical platform to manipulate spin interactions in polariton BEC [20] and superfluid, is the realization of the valley bosonic stimulation, that is the nonlinear amplification of polaritons in specific valleys. The valley bosonic stimulation, however, has remained a major challenge, due to the significant difficulties in generating the polariton ground state densities necessary to induce the stimulated scattering of polaritons in desired valleys. Non-resonant experiments have been studied in which the generation of polaritons is mediated by the population of an exciton reservoir. Yet they usually suffer from the reduced photoluminescence (PL) emission, due to exciton-exciton annihilation processes [21,22] and additional valley depolarization effects, induced by exciton-exciton Coulomb exchange interactions [23,24].

Here, we demonstrate the self-induced valley bosonic stimulation in a $WS_2$ monolayer microcavity. To generate the polariton densities necessary to induce the self-stimulated scattering of polaritons in desired valleys, we resonantly inject polaritons at specific energy and wavevector (k) [25] and with specific spin orientation. Since polaritons are resonantly injected in the microcavity, we do not create an exciton reservoir, thus limiting the density-dependent excitonic effects responsible for the reduction of valley polarization in TMD monolayers [23]. The observed nonlinear behaviour emerges directly from polariton-polariton interactions, and it is not mediated by other type of interactions (e.g., interactions of polaritons with the exciton reservoir and/or the injected photocarriers), which usually dominate the polariton dynamics and nonlinearity in non-resonant excitation experiments [26]. We observe the nonlinear increase of the polariton emission co-polarized with the pump, a key signature of the valley bosonic stimulation. This process is facilitated by the spin-valley locking mechanism inherited by TMD polaritons from their exciton components, allowing polaritons to efficiently retain their valley polarization during their energy and momentum relaxation. Our findings show that polaritons can undergo a self-induced valley-dependent bosonic stimulation, which is the first step toward the realization of spinor polariton BECs in atomically thin TMD monolayers, allowing control and manipulation of spin interactions in polariton lattices and moiré polariton systems for quantum simulations.

The valley bosonic stimulation corresponds to the amplification of polariton emission in specific valleys. The underlying conditions to observe this effect are: (1) the possibility to selectively populate polaritons in specific valley and (2) the preservation of the valley population as polaritons scatter towards the ground state. The spin-valley locking behaviour inherited from TMD excitons (Fig.1a), persists onto the hybrid polariton state, thus ensuring the generation of a ground state polariton population, co-polarized with the pump (condition 1, σ- in Fig.1b). Since the energy and momentum-relaxation of polaritons are spin-dependent [27], spin-down (-up) ground state polaritons stimulate the scattering of other high-energy spin-down (-up) polaritons (condition 2, Fig.1b). Thus, upon increasing the polariton density, the valley bosonic stimulated scattering leads to the nonlinear amplification of the PL intensity and a high degree of valley polarization in momentum space.

The microcavity sample consists of a single monolayer $WS_2$ embedded between two distributed Bragg reflectors (DBRs), grown on top of a Silicon substrate, as schematically shown in Fig.2a. We report the fabrication details and the optical characterization of the microcavity in the supplementary information (SI). In the experiments, the sample is held in a cold finger cryostat at 80 K temperature. To excite the microcavity, we use a visible laser with a central wavelength of 603.2 nm (~2.056 eV). The excitation laser (see SI for details) is spectrally filtered, to obtain pump pulses of 3 meV FWHM, as shown in Fig.2b (red profile). The excitation beam is circularly polarized ($\sigma_+$ or $\sigma_-$) and focused to a 1.7 μm FWHM spot diameter using a 0.60 numerical aperture microscope objective. PL from the lower-polariton branch is collected in reflection geometry through the excitation microscope objective (with ±37° collection angle). The emission from the lower-polariton branch is spectrally filtered, allowing the detection of polaritons populating the bottom of the lower-polariton dispersion at energy ≤ 2.047 eV, while suppressing the excitation laser (Fig. 2B). The PL is then analysed by a polarimeter, composed of λ/4 and λ/2 plates plus a linear polarizer, and projected on the entrance slit of a spectrometer (see SI for details). To excite the microcavity with a specific angle, the excitation beam enters the focusing objective off-centre. We select the specific excitation angle by imaging simultaneously the pump laser and the PL emission, as shown in the SI (S2). The angle and energy of excitation used in the experiments have been estimated in the SI (S3), by considering the scattering relaxation mechanism of polaritons with higher energy, under resonant excitation [14].

Experimentally, the bosonic stimulation regime can be probed by driving the system resonantly with the excitation laser and performing energy and wavevector resolved spectroscopy (Fig. 2b). Specifically, we excite the sample at $E_P$ = 2.056 eV and $\theta_P$ = 11° angle. These excitation parameters allow us to resonantly generate polaritons at $k_P$ = 1.9 $\mu m^{-1}$ close to the inflection point of the lower-polariton dispersion. To filter out the excitation laser, we spectrally filter the emission from the microcavity, as indicated by the green dotted line in Fig. 2b and detect the polariton emission from the lowest energy state of the polariton dispersion. Polaritons injected at $E_P$ and $k_P$ relax down to their ground state, populate the bottom of the lower-polariton dispersion at k~0 and leak out through the top DBR [28]. By repeating the reciprocal space spectroscopic imaging under non-resonant optical excitation (see SI, S4), we collect the lower-polariton dispersion (Fig.S4b) and confirm that the PL emission we detect in our resonant experiments (Fig. 2b) corresponds to polaritons that accumulate in the lowest energy state (at k~0) of the dispersion.

To demonstrate the valley bosonic stimulation of polaritons into the lowest-energy polariton states at k~0, circularly polarized power dependence measurements are performed with selected excitation of $\sigma_+/\sigma_-$ polarized polaritons in both K/K' valleys. The dependences of the intensity (i.e., peak intensity), linewidth (i.e., FWHM), and the central energy (i.e., peak energy) of the polariton emission at different excitation densities were studied by fitting the PL spectra with a single Gaussian function. The PL spectra are extracted as intensity profiles along the y-axis (i.e., energy) at $|k| < 2.6\ \mu m^{-1}$ of the k-space emission data (e.g., Fig. 2b) at different excitation densities. By increasing the circularly polarized ($\sigma_+$) excitation fluence, we increase the polariton density and observe a nonlinear increase of the intensity (Fig.3a, red circles), together with a linewidth narrowing (Fig.3c, red circles), from 4.27 ± 0.02 meV (below threshold) to 4.16 ± 0.02 meV (above threshold), of the co-polarized polaritons emission at the nonlinear photoluminescence threshold. These features are hallmarks of lasing, condensation, and build-up of temporal coherence. We do not observe, however, any obvious blueshift of the peak position with the increasing excitation density (Fig.3e), as in the case of polariton condensates [20].

We also confirm the valley nonlinear behaviour in the K' valley, by exciting the microcavity with the opposite circular polarization ($\sigma_-$) and detecting the co-polarized polaritons. Also in this case, the intensity of the co-polarized polariton emission (Fig.3b, blue circles) increases nonlinearly with the same slope as in the K valley, the linewidth drops (from 4.32 ± 0.02 meV to 4.27 ± 0.02 meV in Fig.3d), while the peak position does not change in energy with the increasing excitation density (Fig.3f). The nonlinear increase of the polariton emission from the lowest energy states is a clear signature of the bosonic final-state stimulation, which is a statistical property of bosons and the driving mechanism behind matter-wave amplification of both atomic [29] and polariton [20] BEC. The bosonic stimulation occurs under less stringent conditions than BEC. The scattering rate of bosons into a certain quantum state is proportional to the number of bosons occupying that state. In particular, the presence of N bosons in the final state enhances the scattering rate of other bosons (by a factor 1+N), thus effectively accelerating the energy relaxation process and stimulating the scattering of other identical bosons in the same state [28]. Therefore, with increasing polariton density, the co-polarized polariton relaxation rate increases, eventually overcoming the threshold for bosonic stimulation. Specifically, this occurs when the relaxation of polaritons towards the ground state overcome its radiative decay, resulting in a super-linear increase of the polariton density at the bottom of the lower-polariton dispersion in both K/K' valley. Since excitons in TMD-monolayers have identical valley-dependent optical response [1], the different thresholds that we observe are due to the generation of different polariton densities in the two valleys, at fixed σ+/σ- excitation power. The latter may be due to the presence of strain on the monolayer, affecting the TMD selection rules [30], and to an unequal excitation of the transverse-electric–transverse-magnetic (TE-TM) polariton modes. At fixed excitation angle and energy, the polariton TE-TM polarization splitting [14], inherited from the cavity [31] and the TMD excitons [32], can result in different σ+/σ- polaritons densities, for an unequal excitation of the TE/TM polariton modes inside the cavity. Since the threshold is ultimately determined by the polariton density in each valley, the valley less populated requires a higher excitation density to reach the bosonic stimulation threshold. However, despite this difference, the nonlinear emission and the linewidth-narrowing that we observe in both valleys, confirm the valley nature of the bosonic stimulation. This is further corroborated by the observation of a self-amplification in both valleys (see SI), reaching a maximum of 1.7, comparable with pump-probe amplifications, observed in similar TMD microcavities, with enhanced polaron-polaritons interactions [33].

Conversely, the cross-circular polarized PL intensities, corresponding to polaritons scattering in the opposite valley respect to the one where they were initially generated (i.e., intervalley scattering), show only a linear increase with increasing fluence and a nearly constant (i.e., within the error bars) linewidth. This confirms that, by resonantly injecting polaritons in the lower-polariton dispersion at specific energy and wavevector, we can generate the polariton density necessary to induce the bosonic stimulation of polaritons in the same valley, thus facilitating the intra-valley scattering of polaritons to overcome the inter-valley scattering, typical of $WS_2$ excitons [34]. In our experiments, no injection of an initial population of polaritons at $k \sim 0$ is required to stimulate the scattering of polaritons in the lowest energy state as, for example, in optical parametric amplification experiments [25]. Similarly, we do not measure any polariton emission at energy and wavevectors above the excitation laser (i.e., “idler” state [25]). We observe a self-induced valley bosonic stimulation in the polariton ground state, confirming that our resonant excitation scheme allows, in each valley, an efficient relaxation of polaritons toward the bottom of their dispersion and the generation of polaritons density sufficient to overcome the stimulation threshold. The bosonic stimulation occurs at pump fluences lower than nonlinear scattering processes in other materials, such as, for example, BBO crystals [35], confirming the high interacting nature of TMD polaritons.

Finally, the spin-valley locking mechanism is uniquely probed by the generation of valley polaritons and the conservation of their spin in each valley (Fig. 4). To investigate the spin conservation of polaritons associated with their valley polarization, we measured the circular degree of polarization (CDP) from the polarization-resolved k-space spectroscopy images acquired during the power dependence measurements. As expected from the optical selection rules, a σ+ (σ-) circularly polarized excitation beam, will generate co-polarized polaritons in K (K’) valley in momentum space (Fig. 4). In other words, resonantly injected polaritons in K (K’) valley highly retain the σ+ (σ-) circular polarization of the pump during their energy and momentum relaxation. The non-unitary circular degree of polarization that we observe is due to additional depolarization mechanisms. Many experimental studies [24,34] have shown that the spin relaxation dynamics in TMDs is dominated by the presence of at least two main depolarization mechanisms: (1) the Maialle-Silve-Sham mechanism [36], which is induced by the longitudinal-transverse splitting of the exciton energy and, ultimately, by the long-range Coulomb exchange interaction between the electrons and holes composing the excitons [37,38]; (2) the intervalley scattering, due to both the long- and short-range exchange interactions [38] and the scattering by acoustic phonons [39,40]. In addition, in tungsten-based materials, the depolarization processes are more pronounced than in other TMD materials, due to the presence of an extra depolarization mechanism, such as the Γ-valley assisted intervalley scattering [34]. Thus, a maximum valley polarization of ~35% has been reported in $WS_2$ monolayers at both cryogenic ($\lesssim$ 90K) [41] and room temperature [41,42]. The combination of the spin-valley locking mechanism of TMD materials and the faster energy relaxation dynamics induced by the valley bosonic stimulation, allows us to observe a maximum degree of valley polarization of 64% for K-valley in $WS_2$ monolayers, which usually displays a low degree of valley polarization as bare monolayer [41,42].

In summary, the following three experimental observations represent a clear indication of the valley bosonic stimulation: 1) the observation of the quadratic nonlinearity in K and K’ valley, 2) the lack of nonlinearity in the opposite valley (respect to where polaritons are generated), 3) the narrowing of the FWHM in both valleys. The first two effects can only be realized if the rate at which polariton scatters toward k~0 state is proportional to its occupation number in each valley. The narrowing of the FWHM, indicating an increase of the coherence time, can only occur for a macroscopically occupied valley polariton mode. Taken together these observations indicate that our resonant excitation scheme allows to overcome TMD depolarization effects [34,37-40], a major obstacle for valley polariton nonlinear studies, suggesting a possible route for realizing strongly spin-polarized polariton condensates and polariton spin lattice for quantum simulations [43] above cryogenic temperature.

In conclusion, we have demonstrated the self-induced valley bosonic stimulation of exciton-polaritons in a monolayer TMD semiconductor. By coherently injecting polaritons at specific energy and wavevector of the lower-polariton dispersion, we overcome the challenges of generating valley polariton densities necessary to explore valley nonlinear processes in TMD microcavities. More importantly, the valley polariton stimulated scattering leads to the unique spin-valley locking nonlinear amplification of the polariton emission, despite the prominent valley depolarization processes typical of TMD materials. The valley-dependent bosonic stimulation offers a new experimental knob for harnessing spin-dependent polariton interactions in the nonlinear regime, paving the way for the realization of spin-polarized polariton BEC in atomically thin TMD monolayers, polariton spin lattices in TMD monolayers and moiré heterostructures, as well as low-threshold spin-polarized lasers.

This work was supported by the Gordon and Betty 365 Moore Foundation (Grant No. 5722) and the Ernest S. Kuh 366 Endowed Chair Professorship.

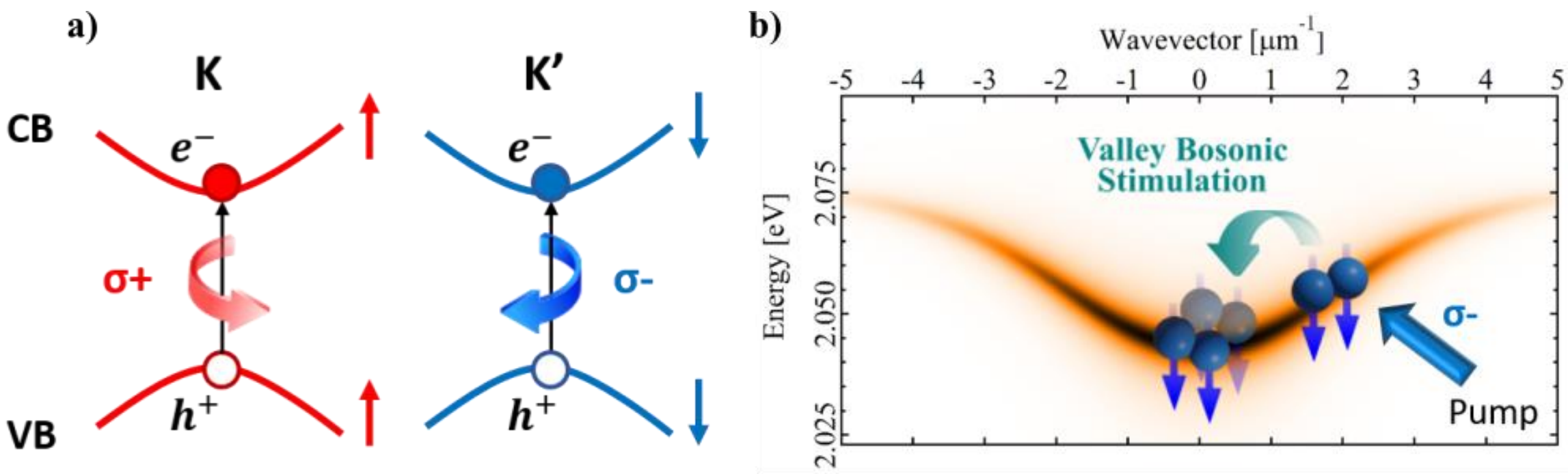

**Fig.1-(a)** Schematic of the spin-valley locking mechanism. Two degenerate valleys form in the $WS_2$ monolayer due to the absence of crystal inversion symmetry. Spin-orbits coupling locks the valley index (K, K') to the electron ($e^-$) and hole ($h^+$) spins (up, down arrows), resulting in specific exciton optical selection rules. Excitons with opposite spins (+1/-1) can be created in different valleys in momentum-space (K/K'), utilizing circular polarized optical excitations (σ+/σ−).**(b)** Schematic of the polariton valley bosonic stimulation induced by the resonant excitation scheme (blue arrow, "Pump"). Polaritons (blue spheres) resonantly injected in the lower polariton branch, with a spin set by the polarization of the pump (σ−), scatter towards the k~0 energy state. The presence of σ− polaritons at k~0, stimulates the scattering of other polaritons with parallel spins, resulting in the self-amplification of polariton emission from the valley-dependent ground state.

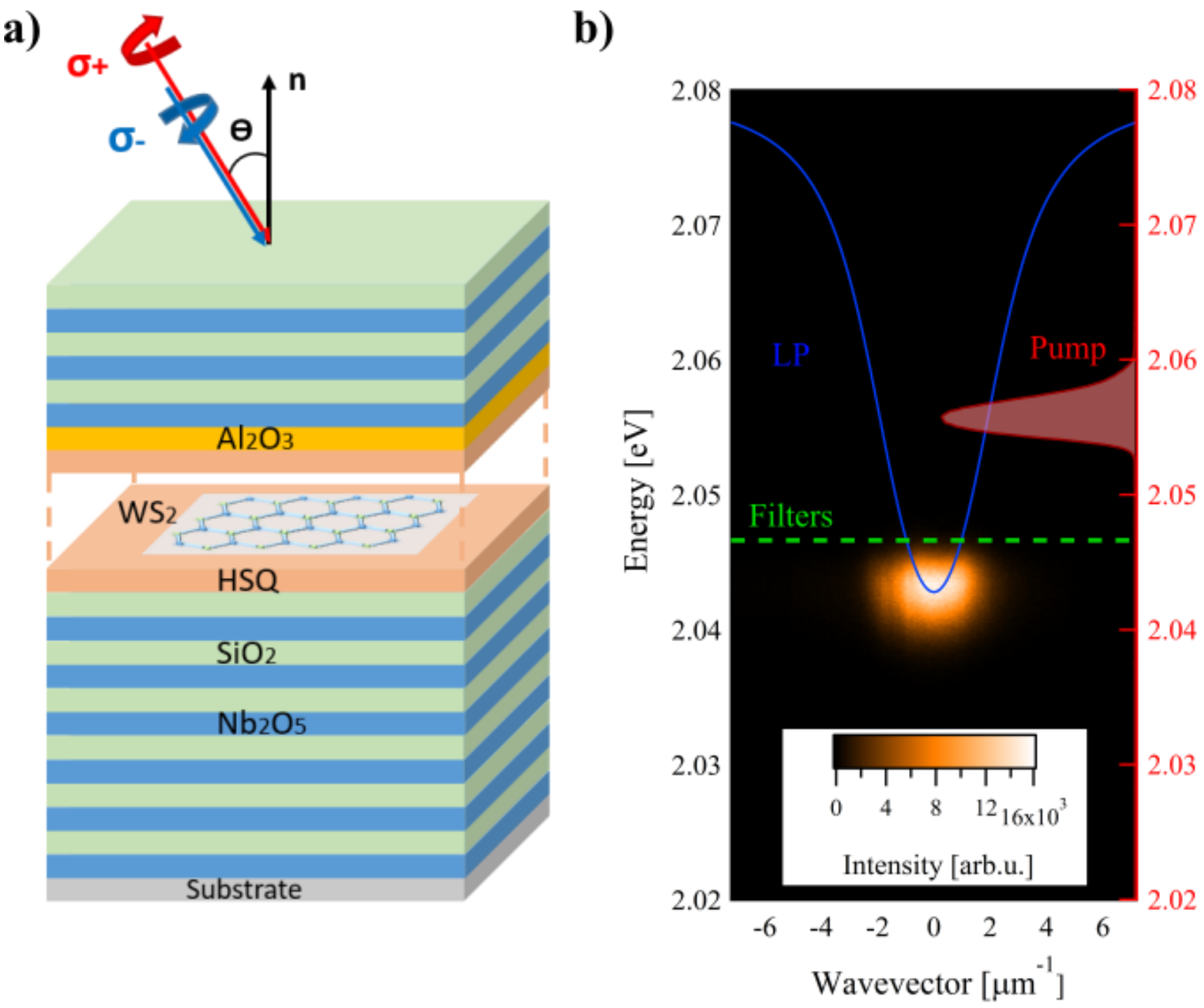

**Fig.2-(a)** Schematic of the microcavity sample, composed of two ($Nb_2O_5/SiO_2$) DBRs and a $WS_2$ monolayer sandwiched between two protective (HSQ and $Al_2O_3$) layers (see SI for details). The microcavity is excited with a $\sigma_+$ ($\sigma_-$) polarized beam set at an angle of incidence ($\Theta$), relative to the microcavity's normal axis (n). **(b)** Experimental energy-wavevector resolved emission intensity measured under resonant σ+-excitation/σ+-detection (137 μJ/cm$^2$ pump fluence). The red trace ("Pump") corresponds to the pump laser intensity profile used in the experiments. To filter out the excitation laser, we spectrally filter the emission from the microcavity sample, as indicated by the green dotted line and detect the PL polariton emission from the ground state of the lower-polariton dispersion.

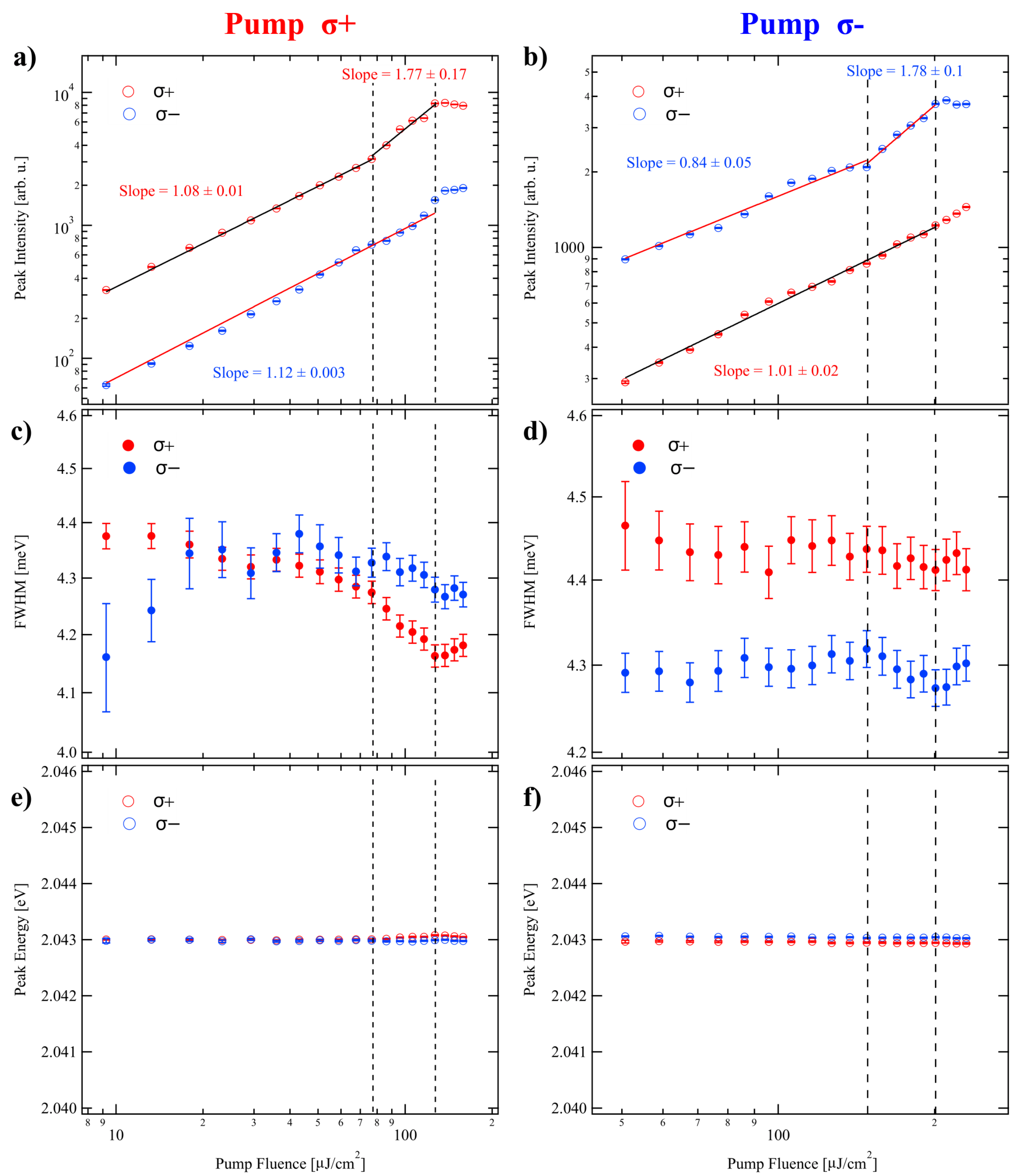

**Fig.3–(a)**The nonlinear increase of the polariton PL intensity co-polarized with the σ+ excitation pump (red circles), showing K-valley polaritons undergo valley bosonic stimulation. **(b)**The nonlinear increase of the polariton PL intensity co-polarized with the σ- excitation pump, confirming the valley bosonic stimulation also in the K' valley (blue circles). In both cases, the cross-polarized PL intensities, $\sigma_-$ in (a) and $\sigma_+$ in (b), exhibit only a linear increase with increasing fluence, as it corresponds to the intervalley scattering of polaritons. **(c)-(f)**With increasing excitation density, the linewidths of polaritons co-polarized with the σ+ **(c)** and σ- **(d)** excitation pump narrow, while the peak energy does not change **(e), (f)**. All the data are plotted on a log-log scale. The data in (a)-(b) are fitted with $y(x) = kx^n$, where n represents the slope of the lines used in the fittings. Error bars are extracted from the fits. For (a),(b),(e),(f) the error bars are too small to be visible. The vertical dotted lines highlight the nonlinear emission regime.

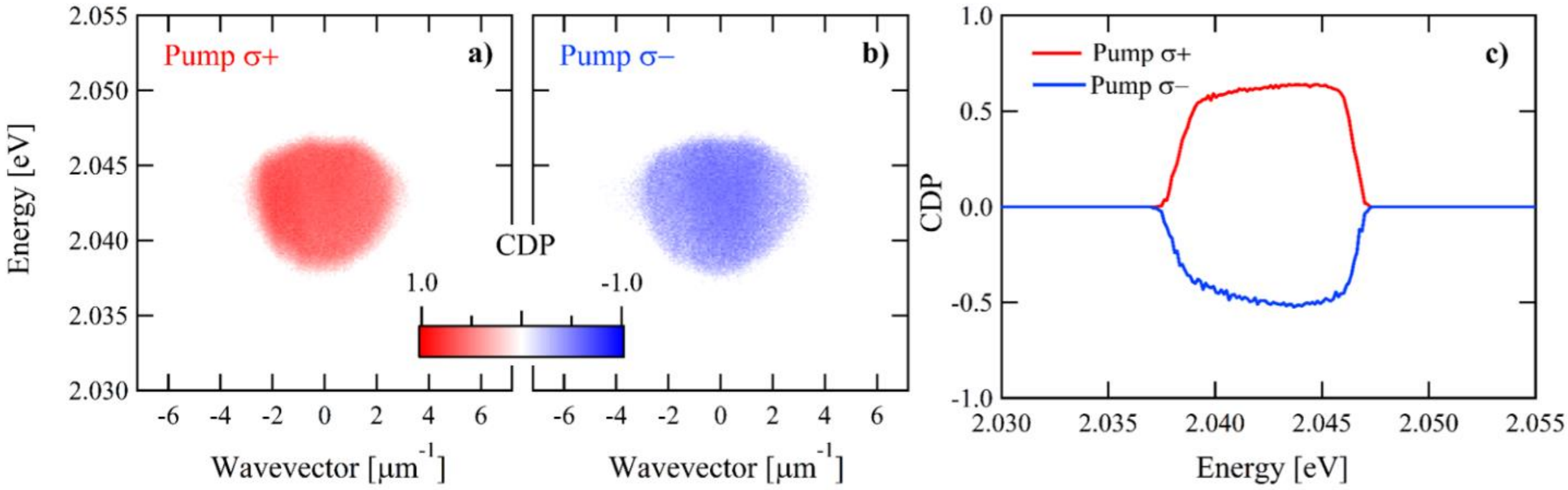

**Fig.4-**The degree of circular polarization (CDP) of the **(a)** K and **(b)** K' valley ground state polariton emission in momentum space. $\sigma_+(\sigma_-)$ polaritons, co-polarized with the pump, scatter toward the bottom of the lower-polariton dispersion and uniformly populate the k~0 energy state in K(K') valley. The σ+(σ-) circular polarization of the pump is largely preserved by K (K') valley polaritons, resulting in a positive (negative) sign of the CDP in each valley. Here, $CDP = (I_{\sigma+} - I_{\sigma-})/(I_{\sigma+} + I_{\sigma-})$, with $I_{\sigma+}$ and $I_{\sigma-}$ being the measured intensities of the two circular polarization components. **(c)** Vertical profiles extracted at $|k| < 1\ \mu m^{-1}$ from (a) and (b).

# Supplementary Information for

# Self-Induced Valley Bosonic Stimulation of Exciton-Polaritons in a Monolayer Semiconductor

Pasquale Cilibrizzi[1], Xiaoze Liu[1], Peiyao Zhang[1], Chenzhe Wang[1],

Quanwei Li[1], Sui Yang[1], Xiang Zhang[1]

[1] NSF Nanoscale Science and Engineering Center, University of California, Berkeley, California 94720, USA

## S1. CHARACTERIZATION OF THE PHOTONIC MICROCAVITY

In this section, we report a more detailed description of the microcavity (MC) sample used in the experiment and the withe light reflectivity measurements, which allow us to measure the photonic dispersion and extract the main parameters of the photonic cavity.

**Description of the microcavity sample.** The microcavity (MC) sample consists of a single monolayer $WS_2$ embedded between two distributed Bragg reflectors (DBRs), grown on top of a silicon substrate, as schematically shown in Fig. 2a of the main text. The bottom DBR consists of 12.5 pairs of $Nb_2O_5/SiO_2$, while the top DBR is composed of 6.5 of the same pairs. The monolayers $WS_2$ flake (HQ Graphene, Inc.) is obtained via mechanical exfoliation and encapsulated between two layers of hydrogen silsesquioxane (HSQ), with the double purpose of protecting the $WS_2$ flake during the fabrication process and preserving its optical and electronic properties. An additional layer of $Al_2O_3$ is deposited on top of the HSQ, to further protect the $WS_2$ flake from the high-temperature plasma during the growth of the top DBR. By controlling the thickness of both HSQ layers, we design the MC to operate at cavity-exciton detuning $\Delta = E_{ph} - E_{ex} = -14$ meV.

**Characterization of the microcavity.** To measure the photonic dispersion, we excite an area of the sample away from the $WS_2$ monolayer and project the Fourier space of the microcavity in the entrance slit of the spectrometer (Fig. S2a).

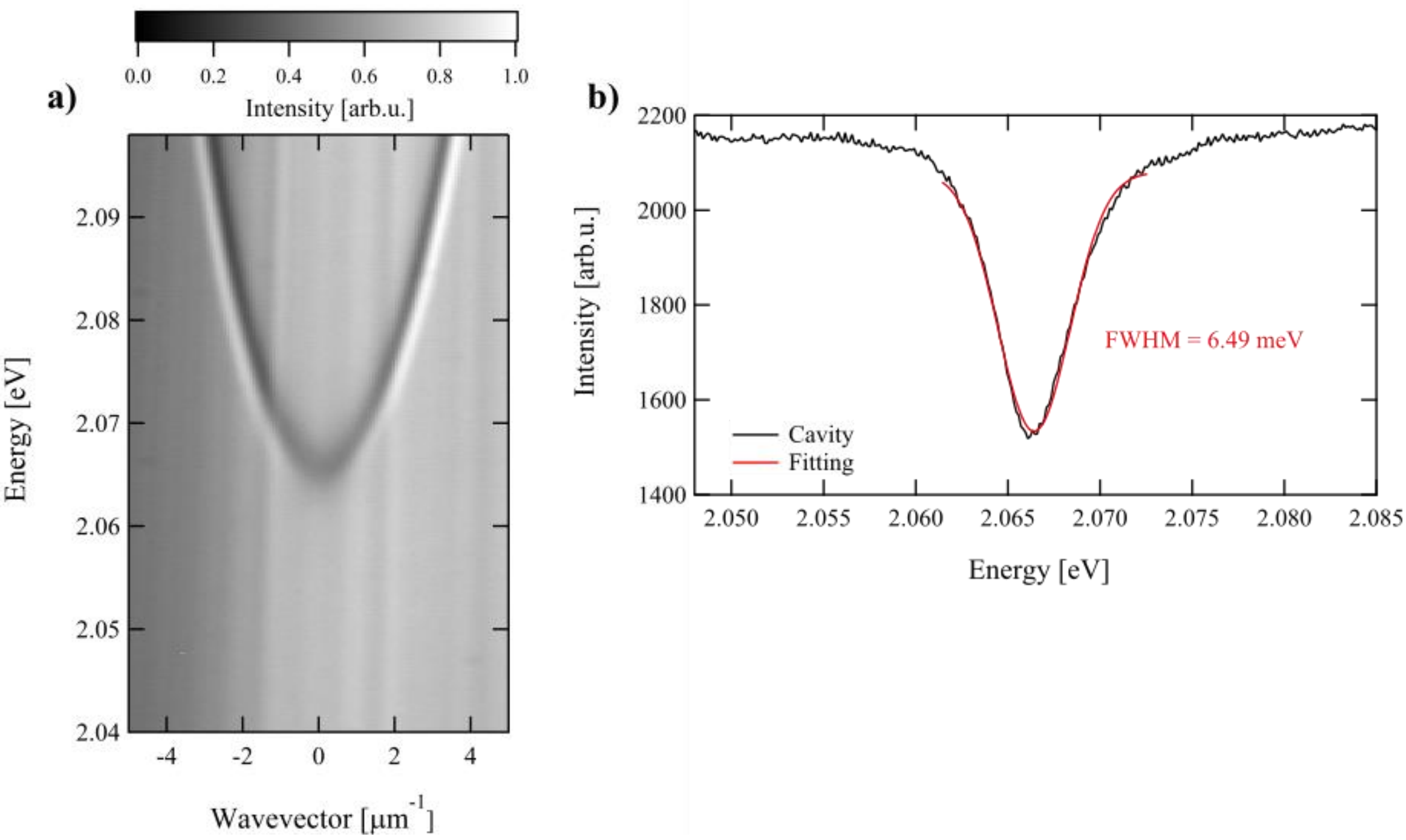

**Figure S2. a)** Experimental white light reflectivity measurements of the MC sample used in the experiment. **b)** Intensity profile extracted along the vertical axis in a), from 2.048eV to 2.085eV and $|k| < 0.5\mu m^{-1}$. The red line represents the inverse Gaussian fitting used to calculate the FWHM and the resonance of the photonic mode of the microcavity.

The main parameters of the photonic MC, extracted from the inverse Gaussian fitting (red line in Fig. S2b) of the bare cavity mode are reported below:

- Bare Cavity Resonance = 2.0664 eV
- Bare Cavity FWHM = 6.49 meV
- Q-factor = 318

## S2. EXCITATION ANGLE

In this section, we report the energy–angle map of the microcavity under resonant excitation. By removing one filter in the detection path (i.e., "detection filters" in Fig. S7), we can collect both the PL and the pump (Figure S3) and confirm the excitation angle in our experiment.

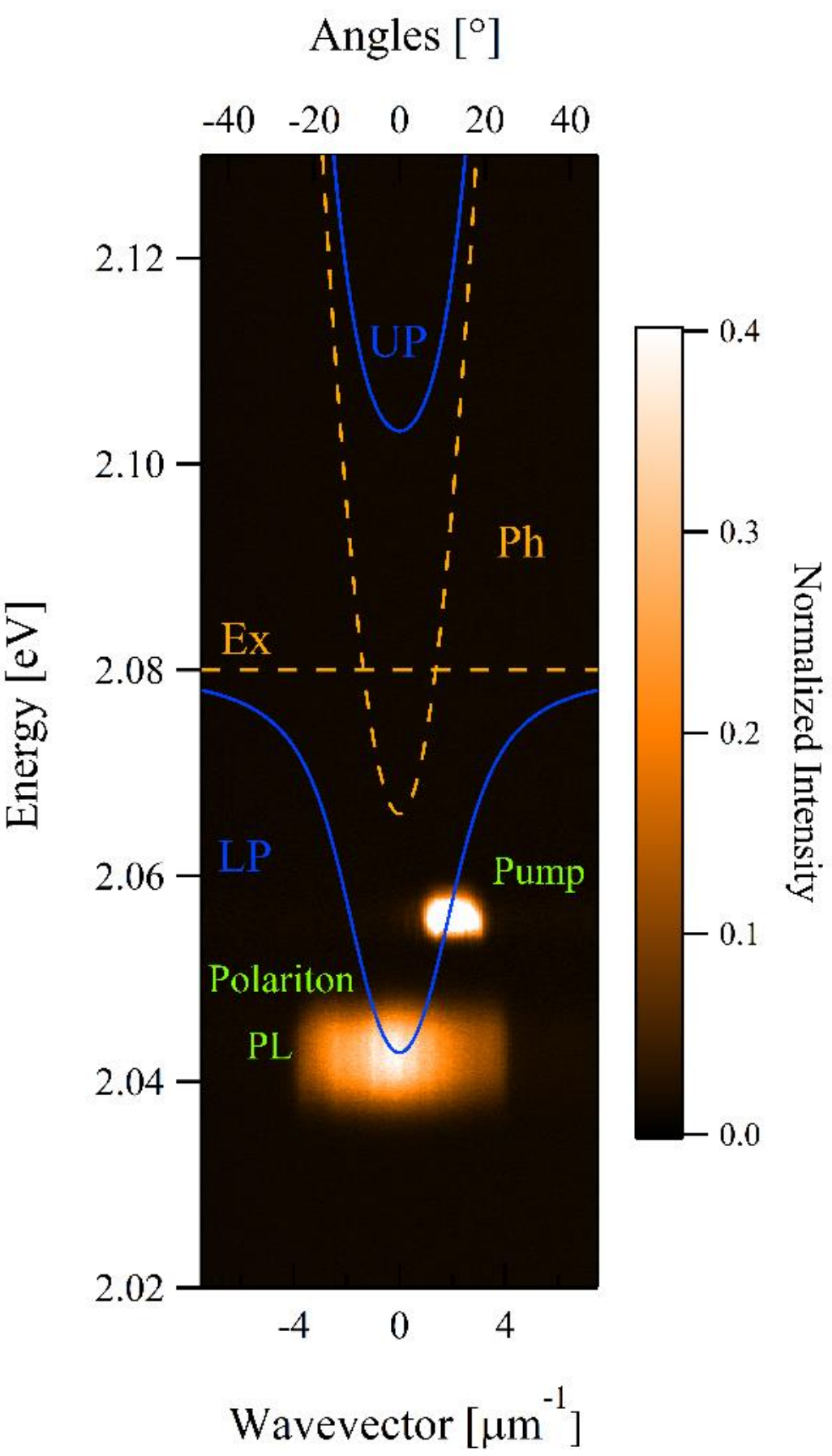

**Figure S3.** Energy and wavevector resolved emission intensity under resonant excitation, showing both the pump laser ("Pump"), which is incident at an angle $\theta_p = 11^\circ$, and the exciton-polariton photoluminescence (PL) emission. The intensity of the pump laser is saturated for illustrative purposes.

The incidence angle is related to the in-plane wavevector ($k_{||}$) through the following relation:

$$k_{||} = \frac{2\pi}{\lambda_p} \sin\theta_p$$

where $\lambda_p$ is the excitation wavelength and $\theta_p$ is the incidence angle of the pump laser. The pump laser is in resonance with the lower polariton branch at an incidence angle of $\vartheta_p = 11°$, with an angular full width at half maximum (FWHM) of ~ 10° and excitation energy of $E_p$ = 2.0557 eV, with FWHM ~ 3 meV.

## S3. ESTIMATION OF THE EXCITATION PARAMETERS

Under resonant excitation, the scattering process of polaritons requires the conservation of both energy and momentum [14]. In particular, two polaritons injected by the resonant pump, with energy and wavevector ($E_p$, $k_p$), scatter into a lower ($E_0$, $k_0$) and higher ($E_H$, $k_H$) energy state, by conserving both energy and momentum. The latter condition requires that:

$$2\, k_p = k_0 + k_H \qquad (1)$$

$$2\, E_p = E_0 + E_H \qquad (2)$$

To estimate the energy and angle of excitation in our resonant experiments, we proceed as follows:

- from the polariton dispersion under non-resonant excitation (Fig. S4b), we extract the highest and lowest occupied energy states ($E_H$, $E_0$ respectively) and their corresponding wavevector ($k_0$, $k_H$) on the lower polariton dispersion (as indicated by the red crosses in Fig. S4b). The experimental values are:

$$E_0 = 2.0428 \text{ eV}, \quad k_0 = 0$$

$$E_H = 2.0686 \text{ eV}, \quad k_H = 3.8\ \mu\text{m}^{-1}$$

- We then apply the conservation of momentum and energy, i.e., equations (1) and (2) above, to estimate the highest energy ($E_p$) and wavevector ($k_p$) of the excitation pump:

From (1): $$k_p = \frac{k_0+k_H}{2} = 1.9\ \mu m^{-1} \qquad (3)$$

From (2): $$E_p = \frac{E_0+E_H}{2} = 2.0557\ eV \qquad (4)$$

The energy and wavevector above can then be converted in their corresponding values of wavelength and angle, by using the following relations (with h and c being respectively the Planck constant and the speed of light):

$$\lambda_p = \frac{h\,c}{E} = \frac{1240\ eV \cdot nm}{2.0557\ eV} = 603.2\ nm \qquad (5)$$

$$\vartheta_p = sin^{-1}\left(\frac{\lambda_p \cdot k_p}{2\pi}\right) = 10.5° \qquad (6)$$

which are the values that we use in our main experiment (Fig.3a and Fig3b of the main text).

In our experiments, the excitation laser is in resonance with the lower polariton branch at an incidence angle of $\vartheta_p = 11°$ and excitation energy of $E_p = 2.557$ eV. However, since the excitation beam covers a finite range of incidence angles (~ 10°), larger than the polariton angular width (~ 7°) at the value of energy corresponding to the energy of excitation ($E_p$), the condition on the energy (eq. 2) is more critical than the condition on the angles or wavevector (eq. 1), as long as the excitation is in resonance with the lower polariton branch.

## S4. POLARITON EMISSION: RESONANT VS NON-RESONANT EXCITATION

In this section, we report the momentum resolved spectroscopy of the MC sample under non-resonant excitation.

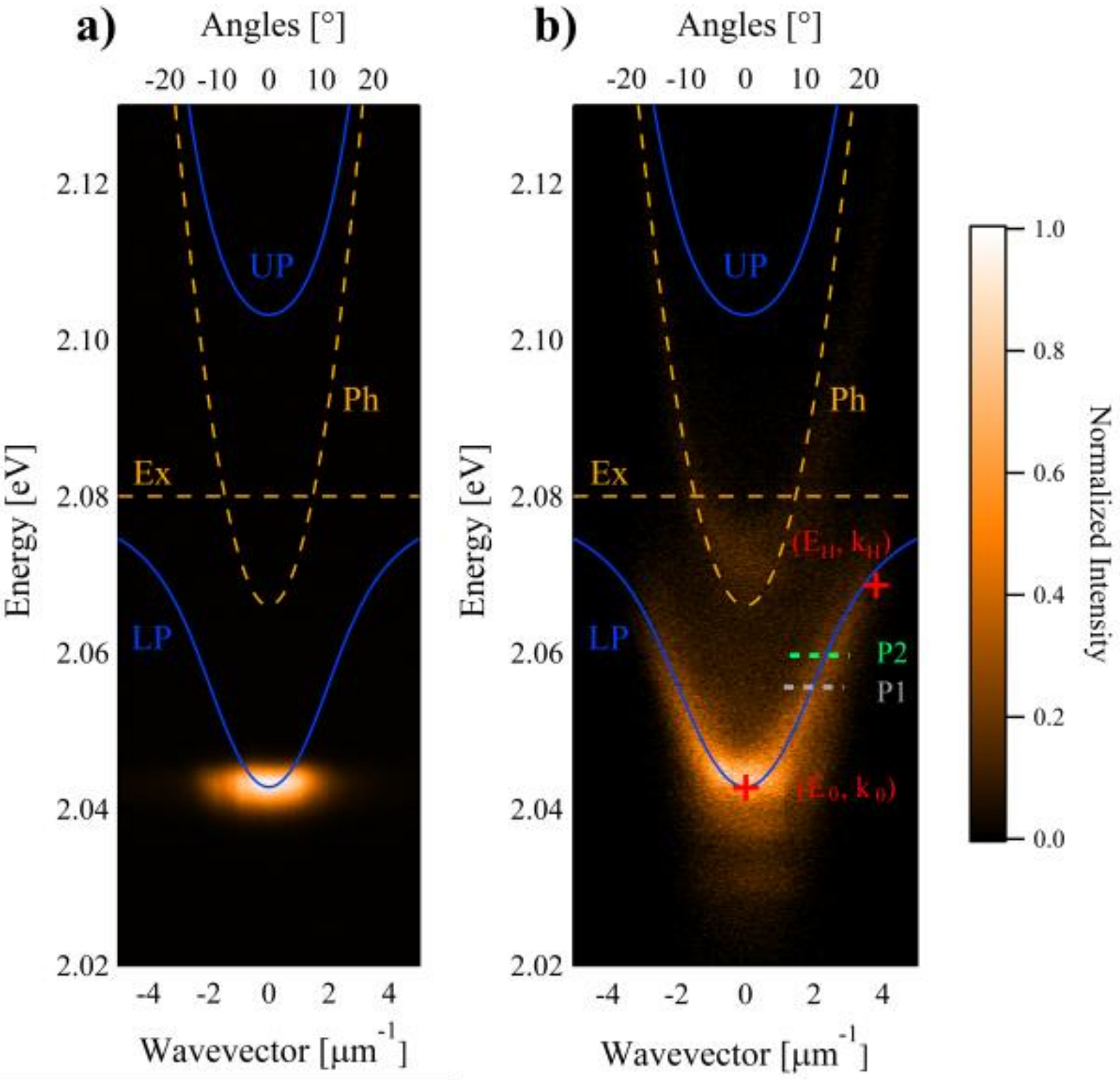

**Figure S4.** Energy and wavevector resolved emission intensity under, **a)** resonant and **b)** nonresonant excitation. The yellow dotted lines represent the theoretical exciton (Ex) and photon (Ph) dispersion, while the blue lines represent the lower (LP) and upper (UP) polariton branches. The red crosses in

b) represent the highest and lowest occupied energy values, that we use to estimate the angle and energy of excitation in section S3. P1 represents the energy of excitation used for the experiment described in the manuscript, while P2 corresponds to the control experiment presented in Section S6. The length of both P1 and P2 dotted lines, correspond to the angular aperture of the pump spot (~10°).

The resonant excitation measurement (Fig. S4a) has been obtained with the experimental parameters reported in the main text, (i.e., $\vartheta_p = 11°$, $E_p = 2.557$ eV). The non-resonant excitation measurement (Fig. S4b) has been obtained by exciting the MC at the first reflectivity minimum above the cavity stopband, i.e., at E = 2.214 eV in reflection geometry. The spectrally and in-plane wavevector (k) resolved emission intensity is shown in Fig. S4b. By comparing the non-resonant dispersion (Fig. S4b) with the measurement performed under resonant excitation (Fig. S4a), we confirm that the PL emission we detect in our experiments corresponds to polaritons that accumulate in the lowest energy state (at k~0) of the lower-polariton dispersion.

## S5. ADDITIONAL DETAILS ON THE POLARITON LINEWIDTHS IN THE NONLINEAR REGIME

In the following we report the values of the linewidths (FWHM) of both co- and cross-polarized components in the nonlinear regime (i.e., corresponding to the FWHM values in region highlighted by the vertical dotted lines in Fig.3c and Fig.3d of the manuscript):

Pump σ+ (Fig. 3c)

| Detection | Experimental Values | Decrease |
|---|---|---|
| σ+ | From 4.27 ± 0.02 meV to 4.16 ± 0.02 meV | 0.11 ± 0.04 meV |
| σ- | From 4.33 ± 0.03 meV to 4.28 ± 0.02 meV | 0.05 ± 0.05 meV |

Pump σ- (Fig. 3d)

| Detection | Experimental Values | Decrease |
|---|---|---|
| σ- | From 4.32 ± 0.02 meV to 4.27 ± 0.02 meV | 0.05 ± 0.04 meV |
| σ+ | From 4.44 ± 0.03 meV to 4.41 ± 0.03 meV | 0.03 ± 0.06 meV |

The decrease is calculated as the difference between the FWHM values at the onset/end of the nonlinear emission regime (indicated by the dotted lines in Fig.3c, d). The linewidths of polaritons co-polarized with the σ+ (Fig. 3c) and σ- (Fig.3d) excitation decreases, while their respective cross-polarized components stay nearly constant (i.e., within the error bars).

## S6. RESONANT EXPERIMENTS AT DIFFERENT EXCITATION ENERGY

In this section, we present circularly polarized (σ+) power dependence measurements performed at excitation angle $\Theta_p$ = 11° and energy $E_p$ = 2.0598 eV. These excitation parameters allow us to resonantly generate polaritons at higher excitation energy (position P2 in Fig. S4b) compared with $E_p$ used to observe the valley bosonic stimulation (position P1 in Fig. S4b).

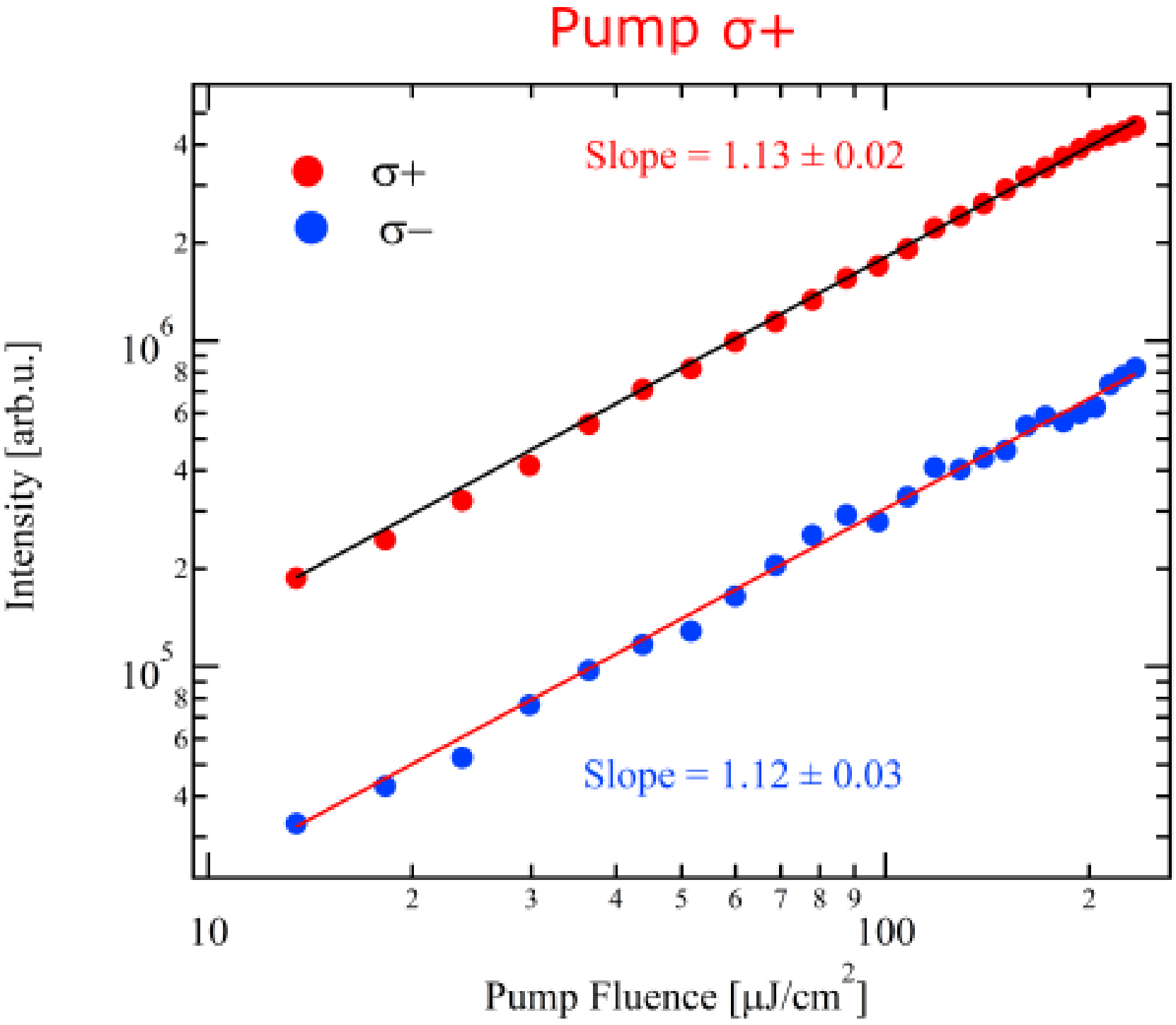

**Figure S5.** Power dependence measurements acquired under circular polarized (σ+) excitation with the same angle (i.e., $\Theta_p$ = 11°) as in Fig.3 of the main text, but with higher excitation energy ($E_p$ = 2.0598eV), as shown by the dotted line P2 in Fig. S4b. The data are plotted in a log-log scale and fit with a power law of the form $y(x) = kx^n$, where n represents the slope of the lines used in the fittings.

We excite the MC with a σ+ polarized excitation beam and detect both the co- (σ+) and the cross-polarized (σ-) polariton emission, shown in Fig. S5 as red and blue trace, respectively. At higher excitation energy, we report a linear increase of the polariton emission in both K and K' valley (Fig.S5), indicating that the condition on the energy (eq. 2 in Sec. S3) is critical for an efficient scattering of polaritons toward the ground state and the consequent observation of the valley bosonic stimulation.

## S7. AMPLIFICATION OF THE POLARITON EMISSION

In this section, we report the PL spectra acquired with σ+ excitation/σ+ detection (Fig. S6a) and σ– excitation/σ− detection (Fig. S6b) at different excitation fluences, corresponding to polariton emission below (blue traces) and above (red traces) the bosonic stimulation threshold shown in Fig.3 of the main text. The PL spectra are extracted as intensity profiles along the y-axis (i.e., energy) at $|k| < 2.6\ \mu m^{-1}$ of the lower-polariton emissions data (e.g., Fig 2b). The amplification ratio, estimated as the ratio of the maximum intensity of the polariton emissions normalized by the excitation fluences below and above threshold, is 1.7 in the K valley (Fig. S6a) and 1.3 in K' valley (Fig. S6b). To calculate the amplification of the polariton emission, the intensity of the emission has been normalized by the excitation density. This excludes that the increase of the polariton emission after threshold is due to the stronger excitation power. The different amplification factors for K and K' valleys are a consequence of the different thresholds that we observe for the opposite circularly polarized excitations. Despite the difference in thresholds, the evidence reported in Fig.3 of the main text, and the amplification greater than one in both valleys, confirm the valley nature of the bosonic stimulation. The maximum self-amplification of 1.7 measured in our experiments is comparable with the amplification of ≥ 2 observed in pump-probe experiments in similar TMD microcavity systems with enhanced polaron-polaritons interactions [33].

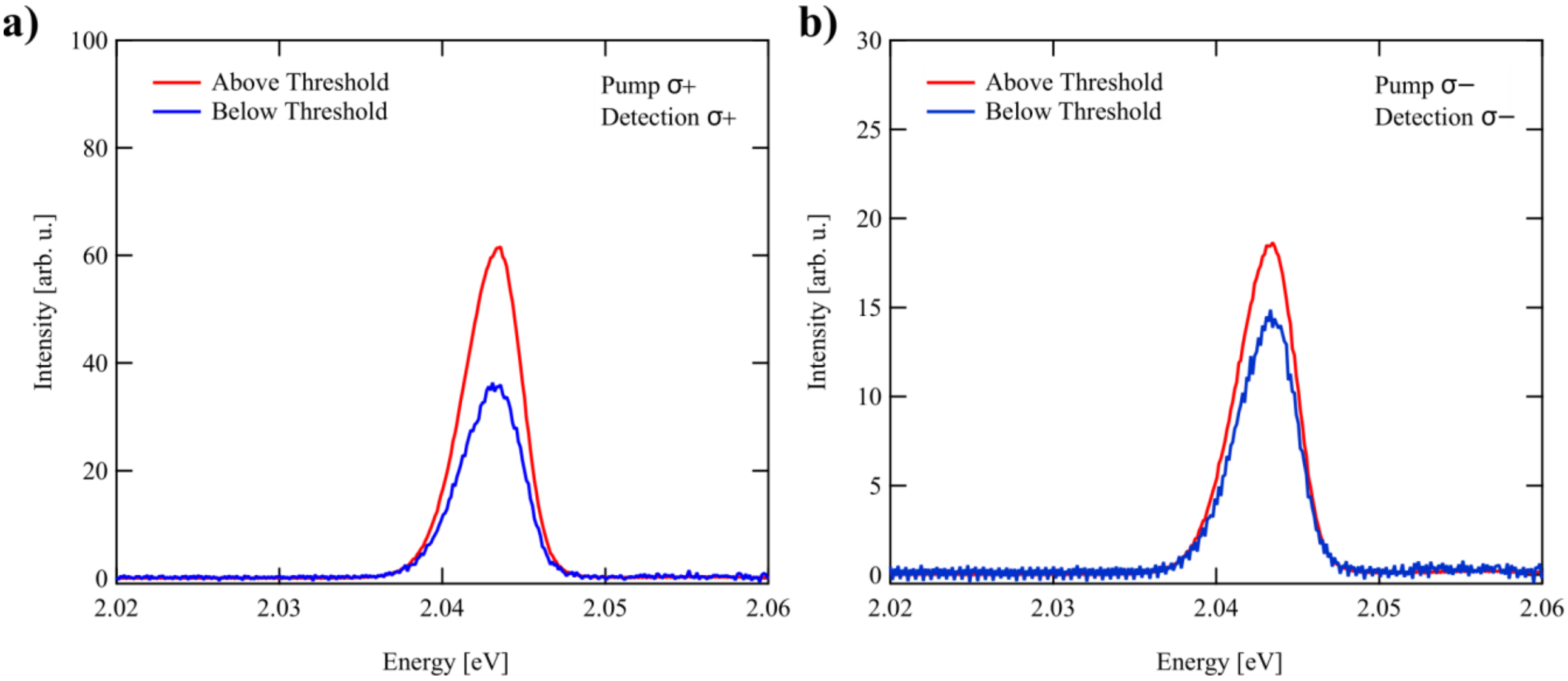

**Figure S6.** Energy resolved intensity profiles of the polariton emission, co-polarized with the excitation pump, in K **(a)** and K' **(b)** valley, normalized by the excitation fluences below (blue trace) and above (red traces) the bosonic stimulation threshold.

## S8. EXPERIMENTAL SETUP

The sample is held in a cold-finger cryostat at a temperature of 80K and is optically pumped at specific energy and wavevector, as indicated in the main text and section S2. This excitation scheme allows us to resonantly pump the lower-polariton dispersion. To excite the microcavity sample, we use a visible laser with a central wavelength of 603.2 nm (~ 2.0557 eV). It is generated via the second harmonic generation of an optical parametric oscillator (OPO) signal beam. The OPO signal is delivered by the Coherent Chameleon Compact OPO (200 fs, 80 MHz) which is pumped by the Coherent Chameleon Ultra II mode-locked Ti-sapphire pulsed laser (140 fs, 80 MHz). The excitation laser is spectrally filtered using a combination of angle-tuned bandpass filters ("excitation filters" in Fig. S7), to obtain pump pulses of ~ 3 meV FWHM, as shown in Fig.2b (red profile) of the main text. The excitation beam is circularly polarized (i.e., σ+ or σ−) and focused to a 1.7 μm FWHM spot diameter using a 0.60 numerical aperture microscope objective. PL from the lower-polariton branch at k~0, is collected in reflection geometry through the excitation microscope objective (ensuring a ±37° collection angle). It is then analysed by a polarimeter composed of λ/4, λ/2 plates, and a linear polarizer (LP) and projected, by means of a lens system, on the entrance slit of a spectrometer.

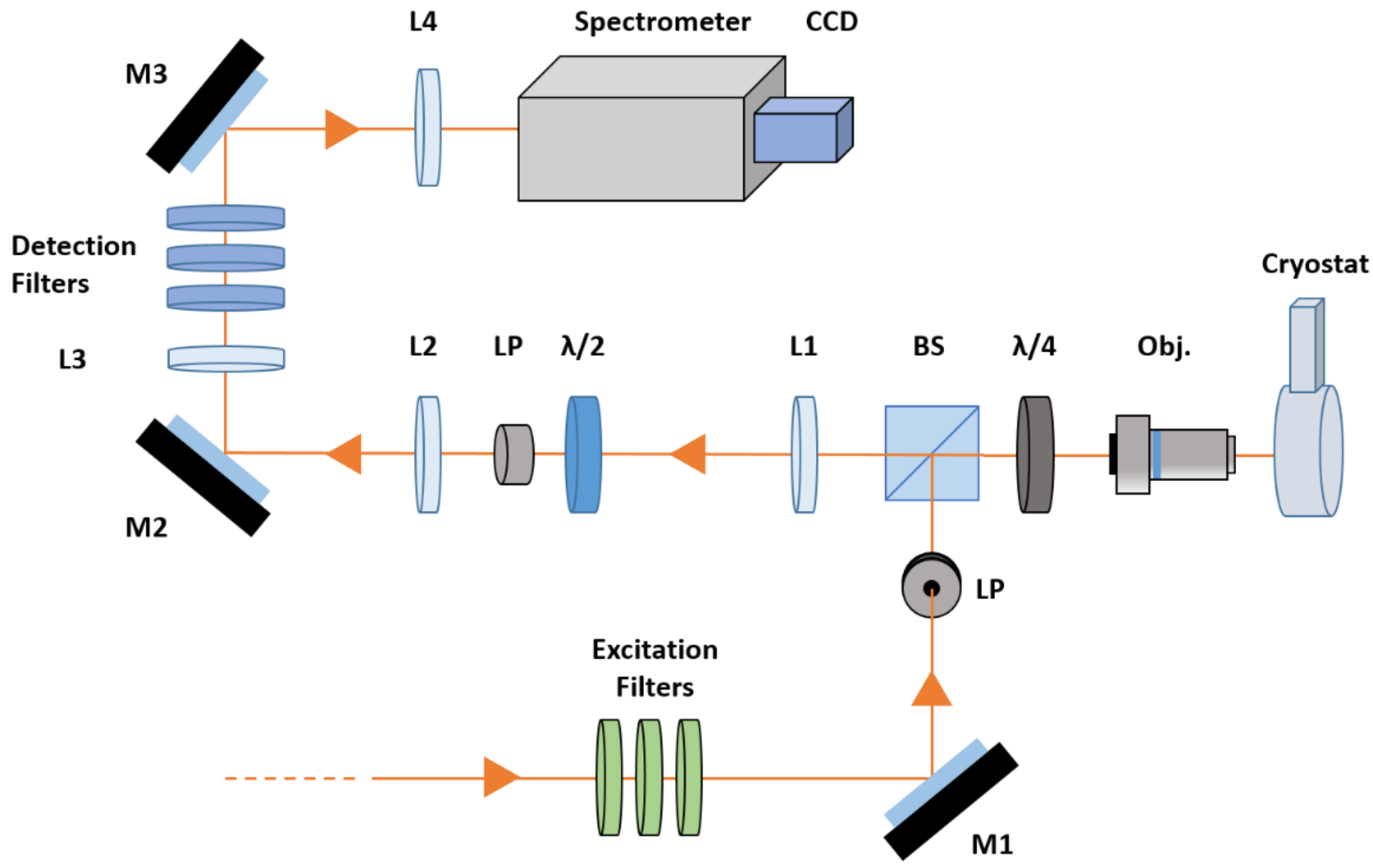

**Figure S7.** Sketch of the optical setup used in the experiments. List of the optical components: **M1, M2,** and **M3** are mirrors; **LP**s are linear polarizers; **BS** is a non-polarizing beam splitter; **λ/4** is a quarter-wave plate; **Obj.** is the objective used for excitation and collection, with 40x magnification and 0.65 numerical aperture; **L1, L2, L3,** and **L4** are convex lenses; **λ/2** and **LP** are respectively the quarter-wave plate and the linear polarizer used to measure the Stokes parameters.

The emission from the lower-polariton branch is then spectrally filtered using another combination of angle-tuned filters ("detection filters" in Fig. S7), to block out the excitation laser and collect only photons with energy $\leq 2.047$ eV. The latter corresponds to the energy range of the bottom of the lower-polariton dispersion. To perform white-light reflection spectroscopy and characterize the MC sample (Fig. S2) we use a broadband (white) light source. A mirror on a flip mount and a second beam splitter was used for the white light to enter the same optical path of the laser. The reflected light is collected by the same objective and projected in the spectrometer by the same combination of lens used for the PL measurements.